\documentclass[aps,iop,twocolumn,amsmath,longbibliography, amssymb,superscriptaddress]{revtex4-1}

\usepackage{dsfont}
\usepackage{bm}
\usepackage[retainorgcmds]{IEEEtrantools}
\usepackage{graphicx}
\usepackage{mathrsfs}
\usepackage{amsmath}
\usepackage{amssymb}
\usepackage{color}
\usepackage{amsfonts}
\usepackage{times,txfonts}
\usepackage{nicefrac}
\usepackage[colorlinks=true,linkcolor=blue,urlcolor=blue,citecolor=blue,pdfusetitle]{hyperref}
\usepackage{soul}
\usepackage[dvipsnames]{xcolor}

\usepackage{verbatim}
\usepackage{placeins}

\newcommand{\ket}[1]{\vert#1\rangle}

\newcommand{\op}[2]{\vert#1\rangle\langle#2\vert}
\newcommand{\expval}[2]{\langle#1\vert#2\vert#1\rangle}
\newcommand{\comm}[2]{\left[#1,#2\right]}
\newcommand{\dagg}[1]{#1^{\dagger}}

\newcommand{\commjb}[1]{{#1}}
\newcommand{\commjbred}[1]{ {#1}}

\begin{document}

	\title{
Optimal quantum control via genetic algorithms for quantum state engineering in driven-resonator mediated networks
	}
	\date{\today}
	\author{Jonathon Brown}
	\affiliation{Centre for Quantum Materials and Technologies, School of Mathematics and Physics, Queen's University, Belfast BT7 1NN, United Kingdom}
	\author{Mauro Paternostro}
	\affiliation{Centre for Quantum Materials and Technologies, School of Mathematics and Physics, Queen's University, Belfast BT7 1NN, United Kingdom}
	\author{Alessandro Ferraro}
	\affiliation{Centre for Quantum Materials and Technologies, School of Mathematics and Physics, Queen's University, Belfast BT7 1NN, United Kingdom}
	\affiliation{Dipartimento di Fisica “Aldo Pontremoli,” Universit\`a degli Studi di Milano, I-20133 Milano, Italy}
	
\begin{abstract}
We employ a machine learning-enabled approach to quantum state engineering based on evolutionary algorithms. In particular, we focus on superconducting platforms and consider a network of qubits -- encoded in the states of artificial atoms with no direct coupling -- interacting via a common single-mode driven microwave resonator. The qubit-resonator  couplings are assumed to be in the resonant regime and tunable in time. A genetic algorithm is used in order to find the functional time-dependence of the couplings that optimise the fidelity between the evolved state and a variety of targets, including three-qubit GHZ and Dicke states and four-qubit graph states. We observe high quantum fidelities (above 0.96 in the worst case setting of a system of effective dimension 96){, fast preparation times,} and resilience to noise, despite the algorithm being trained in the ideal noise-free setting. These results show that the genetic algorithms represent an effective approach to control quantum systems of large dimensions. 		
\end{abstract}

\maketitle 

\section{Introduction}
	
Quantum state engineering is an essential enabling step for a variety of quantum information tasks, including the initialization of quantum simulators \cite{georgescu2014quantum}, the loading of classical data for quantum-enhanced analysis \cite{aaronson2015read}, or the generation of resourceful states in quantum communication networks \cite{muralidharan2016optimal}. In particular, quantum entangled states, which embody a stark departure from classicality, often provide the main resources towards quantum advantage~\cite{horodecki2009quantum}. Therefore the preparation of entangled states in multi-node quantum networks presents a key challenge for realising quantum protocols on near term devices. 

A successful approach towards state and resource generation consists of two steps: \textit{(i)} engineer a suitable time-dependent Hamiltonian with tunable parameters, as allowed by current experimental capabilities in a given physical platform; \textit{(ii)} find and implement proper temporal dependencies (pulse shapes) for these parameters by invoking optimal quantum control techniques~\cite{dalessandro2007introduction, glaser2015training}. This results in tailor-made control schemes pertinent to the specific platform at hand. 

One platform at the forefront of engineering flexible multi-node quantum networks, to which such approach has been successfully applied, is that of superconducting quantum circuits~\cite{clark2008superconducting, kjaergaard2020superconducting, huang2020superconducting, koch2007chargeinsensitive}. So far superconducting architectures have been employed to realise two-qubit gates using frequency-tunable~\cite{dicarlo2009demonstration, barends2014logic, Li2019RealisationOH, barends2019diabatic} and microwave-driven~\cite{chow2011simple, corcoles2013process} artificial atoms. In addition, demonstration of the coupling between artificial atoms and microwave resonators~\cite{majer2007} opened the door for resonator mediated two-qubit gates~\cite{cross2015optimized, puri2016high, paik2016experimental, ghosh2013high} and provided an alternative platform to study cavity quantum electrodynamics~\cite{blais2007quantum} leading to the field of Circuit QED~\cite{blais2020circuit}. Whilst extremely flexible in their design, it has been shown however that operating these superconducting systems with a reduced level control is not only desirable, but necessary in some cases~\cite{falci2017advances, distefano2015population}. Thus finding optimal control protocols that utilize a limited but effective level of control is of practical interest.
	
Optimal control of quantum systems has yielded a range of new methods inspired, in part, by the development of modern machine learning methods. Specifically, neural-network-based reinforcement learning methods~\cite{sutton1998reinforcement, mnih2013playing} have been recognised as useful tools to study quantum systems~\cite{giannelli2022tutorial} in a variety of contexts including state transfer~\cite{paparelle2020digitally, porotti2019coherent, sgroi2020reinforcement, brown2021reinforcement}, quantum thermodynamics~\cite{sgroi2020reinforcement}, circuit architecture search~\cite{kuo2021quantum}, control of dissipative systems~\cite{an2021quantum} and Adaptive Quantum Enhanced Metrology~\cite{palittapongarnpim2017learning}. Reinforcement learning techniques have proven particularly suitable for control problems of increasing dimension when compared to more standard techniques~\cite{zhang2019does}. However, for the most part these techniques cannot be used as closed-loop optimization schemes and therefore are of limited use for optimization on physical quantum systems~\footnote{Noticeable exceptions are discussed in Refs.~\cite{sgroi2020reinforcement, brown2021reinforcement}} and have relatively poor convergence guarantees which necessitate, often expensive, hyper parameter tuning steps. \commjb{This motivates one to consider whether these control problems can be tackled using comparatively less sophisticated techniques that scale well but have greater ease of implementation. With this in mind we consider} evolutionary strategies, which have already been proposed as a scalable substitute to reinforcement learning methods~\cite{salimans2017evolution} and used as an alternative to gradient based parameter updates in both deep reinforcement learning~\cite{such2017deep} and quantum reservoir computing~\cite{ghosh2021realising}. In fact, Natural Evolution Strategies (\textit{NES}) have already been proposed in the context of quantum control~\cite{zahedinejad2014evolutionary,palittapongarnpim2017learning} resulting in the the realisation of fast, high-fidelity single-shot three-qubit gates in frequency tunable superconducting systems~\cite{zahedinejad2015high, zahedinejad2016designing}, while extension to four qubit systems has also been tackled effectively using more standard global optimization techniques~\cite{spiteri2018quantum}. More recently these \textit{NES} techniques have been utilized to address optimal annealling schedules in spin systems~\cite{hegde2022genetic} and optimal transport of Majorana fermions in superconducting nanowires~\cite{coopmans2021protocol}. The aim of this work is to investigate the potential of these techniques in the specific context of direct state preparation in a driven-resonator mediated, reduced-control multi-qubit network.

Here we consider a register of qubits coupled \commjb{via a common resonator and operated in a regime which facilitates a reduced level of control}, and employ a genetic algorithm to find optimal pulse sequences to drive their dynamics. In order to illustrate our approach, we present efficient control schemes for preparing entangled three- and four-qubit states, including GHZ, Dicke, and graph states, and assess performance against relevant decoherence sources finding the thresholds that limit the quality of our results. The generation of high-quality states{, in short times compared to typical multi-qubit circuit timescales,} is thus demonstrated while identifying fully the sequence of driving pulses to use. Our approach contributes to the growing argument that a hybrid take to quantum control -- that mixes machine learning and optimal control -- is a viable route to the engineering of crucial resources for quantum information processing.

The remainder of the manuscript is organised as follows. In Sec.~\ref{sys_sec} we present the specifics of the system considered and formalise the Hamiltonian. In Sec.~\ref{CGA}, we present an overview of the Continuous Genetic Algorithm employed then follow by formalising the algorithm for the problem of quantum control in section~\ref{CGAOC}. Sec.~\ref{res_sec} focuses on the presentation of the resulting control schemes for the preparation of three-qubit GHZ and Dicke states as well as a specific instance of a four-qubit graph states. In Sec.~\ref{decoherence}, the effect of decoherence is investigated. Sec.~\ref{conc} offers our closing remarks and a forward look.
	
\section{System}\label{sys_sec}
We consider a system composed of $N$ identical and non-interacting qubits coupled via a common single-mode driven resonator. We model such system with the Hamiltonian $H = H_0 + H_{int} + H_d$, where
	\begin{align}
		&H_0= \omega_c a^{\dagger}a + \sum_{j=1}^{N}\omega_j \sigma_j^{+} \sigma_j^{-},\\
		&H_{int} =\sum_{j=1}^{N} g_j (a^{\dagger}\sigma_{j}^{-} + a\sigma_{j}^{+}), \\
		&H_d = \xi (a e^{i \omega_d t} + a^{\dagger} e^{-i \omega_d t} ).
	\end{align}
In writing these Hamiltonians, we have assumed the rotating wave approximation and used units such that $\hbar  =1$. Here $a$ denotes the annihilation operator of the resonator mode, whereas $\sigma_j^{+}=|1\rangle\langle0|$ and $\sigma_j^{-}=|0\rangle\langle1|$ are the raising and lowering operators of the $j$-th qubit. Moreover, $\omega_c$ denotes the resonator frequency, $\omega_j$ the transition frequency of the $j$-th qubit and $g_j$ its coupling strength with the resonator. The driving amplitude (assumed to be real) and the carrier frequency of the drive are indicated with $\xi$ and $\omega_d$ respectively.

The use of a resonator to mediate two-qubit gate interactions is well studied. For example, in the context of superconducting systems the Resonator-Induced Phase (RIP) gate ~\cite{puri2016high, cross2015optimized, paik2016experimental} utilizes two qubits dispersively coupled to a common microwave resonator. In such a \textit{dispersive} regime, the coherent qubit-resonator interaction term becomes negligible leading to an effective qubit-qubit interaction term. \commjb{Whilst this can be beneficial for protection against decoherence there are two main drawbacks: 1) The dispersive coupling precludes real photon processes between the resonator and each qubit which necessitates the use of local qubit drives in order to introduce energy into the system. This subsequently makes this regime adverse to scaling to larger numbers of qubits. 2) The large detunings involved result in small effective qubit-qubit coupling strengths at the expense of gate operation time}. On the other end of the spectrum, \textit{resonant} regimes have been considered to selectively tune qubits in and out of coupling with a common resonator~\cite{haack2010resonant}. In addition, external microwave drives are also commonly used such as with the the cross-resonance gate~\cite{chow2011simple, corcoles2013process}. In contrast to such previous works, we assume in $H$ a fully resonant regime which allows us to adapt a limited control scheme, in that the local qubit drives can be replaced by a single resonator driving term where the amplitude of this drive and the qubit-resonator couplings are tunable. \commjbred{This facilitates the possibility of multi-qubit interactions that operate on timescales that are much shorter than typical multi-qubit gate times.} 
	 
	 Taking the interaction picture with respect to the frequency of the harmonic mode and using $\tilde{H} =  e^{iRt}He^{-iRt} - R$, where $R=\omega_c \left(\dagg{a} a + \sum_{j=1}^{N} \sigma_{j}^{+}\sigma_{j}^{-}\right)$, we get
	\begin{equation}
	    \label{detuning_ham}
		\tilde{H} =\sum_{j=1}^{N}\left[ \delta_j \sigma_j^{+} \sigma_j^{-} + g_j (a^{\dagger}\sigma_{j}^{-} + a\sigma_{j}^{+})\right] + \xi (a e^{i \delta_d t} + a^{\dagger} e^{-i \delta_d t} ),
	\end{equation}
where $\delta_j = \omega_j - \omega_c$ and $\delta_d = \omega_d - \omega_c$ are the detunings between the $j^\text{th}$ qubit and the harmonic mode, and between the drive frequency and the harmonic mode respectively. The utility of this transformation becomes apparent when one assumes full resonance between the qubit transition frequencies, drive frequency and resonator frequency -- namely, $\omega_c = \omega_d = \omega_j$, $ \forall j =1,...,N$. The Hamiltonian thus takes  the simpler form
	\begin{equation}
		\tilde{H} =\sum_{j=1}^{N}g_j(t) (a^{\dagger}\sigma_{j}^{-} + a\sigma_{j}^{+}) + \xi(t) (a  + a^{\dagger}),
		\label{Htilde}
	\end{equation}
	which comprises $N+1$ terms embodying an equal number of controls, where we assume each of the qubit-cavity couplings $g_j (t)$ and the drive amplitude $\xi(t)$ to be time-dependent controllable parameters.
	
	The above Hamiltonian governs the dynamics of the system via the time-dependent Schrödinger equation. Thus if the system is initially in some state $\ket{\psi(t_0)}$, then the time-evolved state of the system at any future time $t>t_0$ is given by 	\footnote{An interesting consequence of the particular form of $\tilde{H}$ in Eq.~(\ref{Htilde}) is that, if the system is initialized in a pure state with real coefficients, then the time-evolved reduced state of the qubit will always be described by real coefficients. This is of course a restriction if one wishes to prepare more general states, however it simplifies the optimization when considering completely real target states. One can ease this constraint if necessary by introducing non-zero detunings in Eq.~(\ref{detuning_ham})} 
	
	\begin{equation}
		\ket{\psi(t)} = \mathcal{T} e^{i\int_{t_0}^{t} \tilde{H}(t') dt'}\ket{\psi(t_0)}.
		\label{UnitaryEv}
	\end{equation}
The goal here is to find the optimal functional forms for $g_j (t)$ and $\xi(t)$, such that the system is dynamically steered in some desired way. Specifically, we are interested in state preparation within the qubit subspace, so we first determine the reduced state of the qubit network
	\begin{align}
		\rho_{Q}(t) =\text{Tr}_{c}\Big(\op{\psi(t)}{\psi(t)}\Big) = \sum_{i=0}^{\infty} {}_c\langle{i} \op{\psi(t)}{\psi(t)} {i}\rangle_c,
		\label{ReducedQubit}
	\end{align}
	and work to find $g_j (t)$ and $\xi(t)$ such that the fidelity
	\begin{equation}
		\mathcal{F}(\rho_{Q}, \sigma) =\langle\psi_\sigma\vert\rho_Q\ket{\psi_\sigma} 
		\label{Fidelity}
	\end{equation}
	is maximized, where $\ket{\psi_\sigma}$ is the target state of interest. It is worth stressing that our approach would work equally -- {\it mutatis mutandis} -- with mixed target states. 

\section{The Continuous Genetic Algorithm}
\label{CGA}
Evolutionary strategies are a class of direct search optimization techniques, drawing inspiration from Darwinian evolution, that have recently been proposed as a viable substitute for gradient based parameter optimization in Neural Networks~\cite{such2017deep} and quantum reservoirs~\cite{ghosh2021realising}, as well as a scalable alternative to Reinforcement learning techniques~\cite{salimans2017evolution}. Of particular interest to continuous-control problems is the so-called ``Continuous Genetic Algorithm" (CGA)~\cite{haupt2004practical}, which generalises the more traditional Discrete Genetic algorithm to allow for continuous parameter values. 

Consider an optimization problem with $\mathcal{N}_{var}$ parameter variables $p_i$. We call a specific instance of these parameters, $\mathcal{C} = [p_1, p_2,..., p_{\mathcal{N}_{var}}]$, a \textit{Chromosome}. This embodies one proposed solution to the optimization problem. One then defines a \textit{Fitness function}, $f(\mathcal{C})=f({p_1, p_2,...,p_{N_{var}}})$: $\mathbb{R}^{N_{var}} \to  \mathbb{R}$.  This function will be determined by the optimization task under consideration and will assign a numerical score to each proposed solution (chromosome) depending on its usefulness (fitness) with respect to the specified task at hand. In the name of generality we assert that $p_i \in \left[-1,1\right]$, where the parameter values are suitably scaled within the fitness function. We call each new iteration of the algorithm a \textit{Generation} where the algorithm proposes several chromosomes to make up a \textit{Population}. The algorithm can be qualitatively summarised using the following steps:

\begin{enumerate}
	\item \textit{Initialization.} Define the fitness function (determined by the optimization task) and fix the hyper-parameters for the genetic algorithm. One then generates an initial population with $\mathcal{N}_{pop}$ chromosomes, typically randomly, which acts as the zero-th generation of the algorithm.
	\item \textit{Natural Selection.} The \textit{fitness} of each chromosome is assessed with a call to the fitness function. The population is then ranked based on these fitness values and the $\mathcal{N}_{survive}$ highest scoring chromosomes are chosen to survive, while the rest are discarded to be replaced by new \textit{offspring} chromosomes in the next generation.
	\item \textit{Pairing.} From the $\mathcal{N}_{survive}$ surviving chromosomes chose $\mathcal{N}_{parents}$ pairs of parent chromosomes in order to produce $(\mathcal{N}_{pop} - \mathcal{N}_{survive})$ offspring to repopulate. Parent chromosomes are sampled probabilistically based on their relative fitness, such that the fittest chromosomes reproduce more frequently, where cloning (using the same chromosome for both parents) is disallowed.
	\item \textit{Mating.} Here, each pair of parent chromosomes is combined in some manner to produce enough offspring to replenish the population. In the simplest case, one can consider cutting the parent chromosomes in half and producing two offspring, made up of the two possible combinations of these halves. Alternatively, one could pick indices along the chromosomes at random, and generate two offspring by first copying the parent chromosomes, then swapping the parameter values corresponding to these indices between the two copies. {This direct transfer of parameter values to generate offspring is the simplest and most obvious way of mating two parent chromosomes. However this approach simply provides offspring chromosomes made up entirely of parameter values that were already present in the previous generation and, as such, introduces no new ``genetic material" in the form of novel parameter values for a given parameter. To tackle this we can implement a combination of the form
	\begin{equation}
	\label{combination}
		p_{i, \textit{ offspring}} = \beta  p_{i,\textit{ parent 1}} + (1-\beta)  p_{i,\textit{ parent 2}}.
	\end{equation}
	This allows us to introduce an element of exploration by allowing not only parameter values present in the previous generation but also any continuous value in between}, modulated by random variable {$\beta \in [0,1]$}.
	\item \textit{Mutation.} This final step introduces further exploration into the search by choosing, at random, a number of elements within each chromosome to be replaced with a new random value. The rate of this mutation is set by a parameter $0 < \alpha < 1$, which determines the number of indices to be targeted relative to $\mathcal{N}_{var}$. This mutation step is applied to the entire population except for the {single} chromosome with the highest fitness, which is known as \textit{Elitism}. This is important to ensure theoretical guarantees of convergence. Specifically it ensures that the maximum achieved fitness is always at least maintained in new generations.
\end{enumerate}
\vspace*{0.5pt}
The algorithm repeats steps 2-5 until convergence or an acceptable level of fitness has been achieved. \commjb{A key point to note here is that the CGA is a stochastic optimization process so with finite time the convergence to any absolute optimal strategy is never guaranteed, however on average one would expect an increase in performance as the computational time grows.} The aforementioned hyper-parameters  associated with CGA then are: population size $\mathcal{N}_{pop}$,  number of survivors $\mathcal{N}_{survive}$ to keep in each iteration, number of parental pairs to mate $\mathcal{N}_{parents}$ and the mutation rate $\alpha$. Also, as discussed above, we have some freedom in how we implement the mating procedure.

\section{Continuous Genetic Algorithms for Optimal Quantum Control}
\label{CGAOC}
In order to apply the CGA we first must formulate the optimal control problem in a suitable manner. As is common practice, we discretize the evolution time into $T$ time intervals of equal duration $\tau$ --- which is manually chosen --- and assume the functional form for each control to be defined by its values at the $T+1$ times $t = 0,\tau, 2\tau,...,T\tau$ (connecting the latter with a simple tanh function, similar to the use of piecewise error functions~\cite{zahedinejad2015high,zahedinejad2016designing}. See Appendix~\ref{app1} for specifics). Therefore, given that each control function is completely defined by these $T+1$ values, and there are $N+1$ controls in total as per Eq.~(\ref{Htilde}), then the total number of parameters that we have to optimize over is $\mathcal{N}_{var} = (N+1)(T+1)$ (which as said is the length of the chromosomes). We then need to outline how we asses the fitness of each chromosome and in doing so define the optimization task.

\begin{figure}[t!]
	\begin{minipage}{\linewidth}
		\includegraphics[width=1\columnwidth]{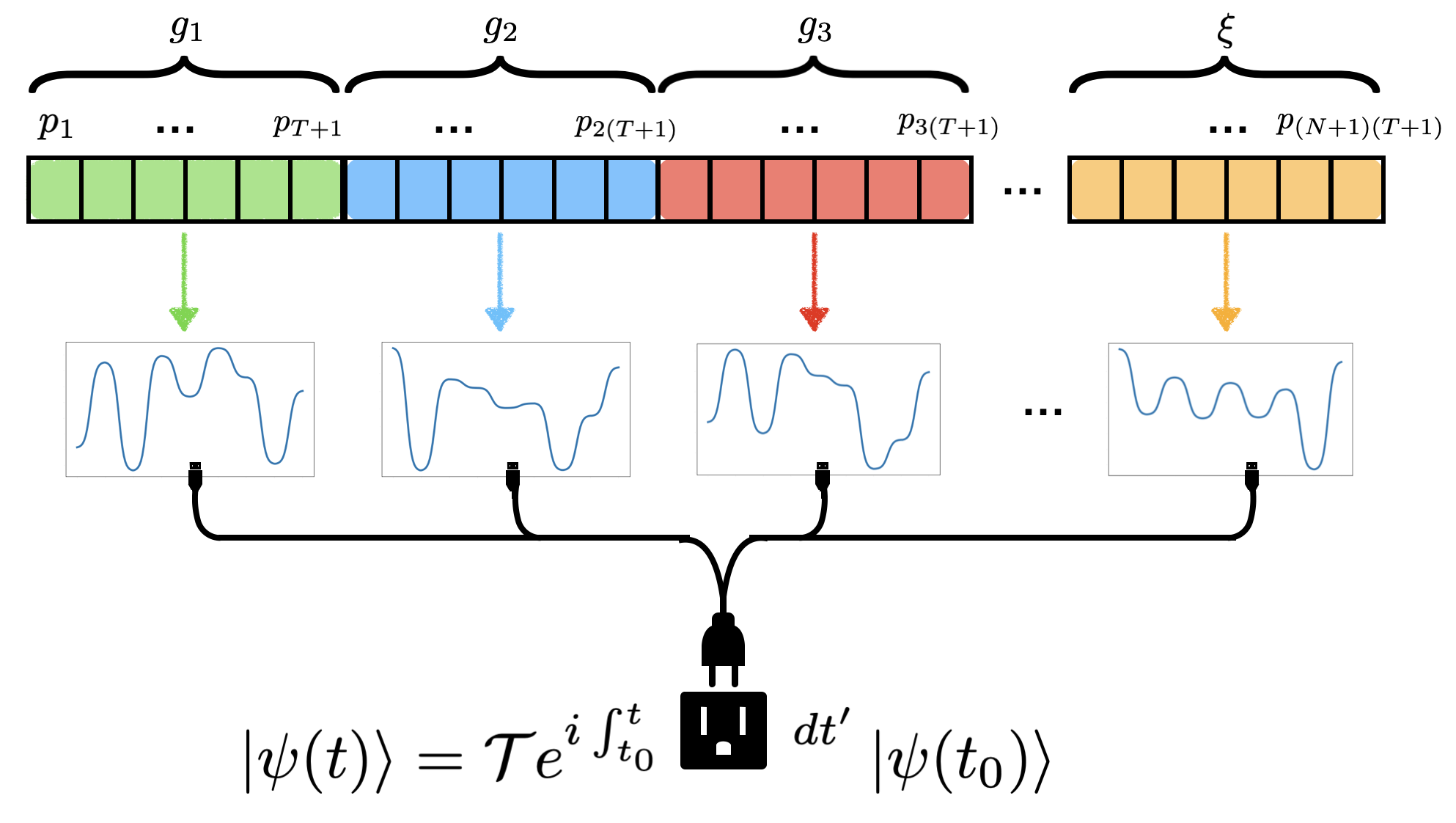}
		\caption{Schematic representation of how the variables within each chromosome (visually represented by the sequence of connected squares) are assigned to which control pulse, depicted along the top of the diagram. Along the bottom shows sample control pulses generated using random parameter values and the function construction method outlined in Appendix~\ref{app1}.}
		\label{ChromeSchematic}
	\end{minipage}
\end{figure}

\subsection{The Fitness Function}
As said, each chromosome is a sequence of $\mathcal{N}_{var}$ parameter values, and there are $N+1$ control pulses. Therefore we first break each chromosome up into $N+1$ sequences where the first $T+1$ parameters determine the first control pulse and so on, as in Fig.~\ref{ChromeSchematic}. Then the parameter values corresponding to each control are scaled accordingly where each of the qubit-resonator couplings assume a common range $g_j \in[-g_0, g_0]$, and the drive amplitude is in the range $\xi \in [-\xi_0, \xi_0]$.  The scaled parameter values are then used to build the control pulses as in Appendix~\ref{app1}, which fully define the time dependent Hamiltonian $\tilde{H}(t)$ in Eq.~\ref{Htilde}. After defining an initial state $\ket{\psi(t=0)}$ the system evolves unitarily according to $\tilde{H}(t)$  (as in Eq.~\ref{UnitaryEv}), where we keep track of the state of the system at every intermediate time between $t=0$ and $t=\tau T$. For each state in this history of states, we trace out the cavity system to obtain a state history for the qubit subspace, $\rho_{Q}(t)$. Next we calculate $\mathcal{F}(\rho_{Q}(t), \sigma)$, which is the time dependent fidelity induced by the specific control pulses. In order to assign a numerical fitness value to the chromosome we simply take the maximum fidelity achieved in the qubit subspace throughout the induced dynamics, i.e
\begin{equation}
	\textit{Chromosome fitness} = \max_{t}\left[ \mathcal{F}(\rho_{Q}(t), \sigma)\right]. 
	\label{Fitness Func}
\end{equation}
In fact, the function actually used is a slight variation of that presented above necessitated by the specific details of the simulation and the ability to ``extract" the state with the highest fidelity (see Appendix \ref{simulation_details} for details). 

\section{Results}
\label{res_sec}

\subsection{Analysis of performance: case studies}

Below we outline the optimal control schemes found when applying the CGA approach to prepare genuinely entangled 3 and 4 -qubit states from completely separable initial states. In doing so this acts as a proof of principle --- for both the Hamiltonian $\tilde{H}$, and the optimization method --- with respect to entanglement generation in general and state preparation specifically. Below we consider the physically reasonable maximum coupling and driving $g_0, \xi_0 = 2\pi~ 200$MHz, and total time-scales $\tau T \leq 10$ns (see appendix \ref{algo_details} for full details). \commjbred{An important point to make is that such timescales are considerably shorter than the typical gate times associated with two-qubit operations on superconducting platforms, which are commonly of the order of $100$ns {\cite{kjaergaard2020superconducting, huang2020superconducting}}. This means that each of the following protocols operate on a much faster timescale than their circuit counterparts which prepare the same state.} We specifically consider 3 states: GHZ state, three-qubit Dicke states with two excitation, and the 4 qubit ``Box" Cluster state. 

\paragraph{GHZ state.} GHZ states are relevant genuinely tripartite entangled states~\cite{dur2000three} defined by
\begin{equation}
	\ket{GHZ} = \frac{\ket{000} + \ket{111}}{\sqrt{2}},
\end{equation}
We assume the system to initially be in the state $\ket{\psi_0} = \ket{010}_Q \ket{0}_{cav}$. We use $\ket{010}$ as the initial state of the qubit network in this specific instance, as opposed to simply the global vacuum, since the latter has non-zero fidelity to the target GHZ state and thus starts the optimization in at undesirable local maxima. Such initial preparation is not so restrictive given that single qubit rotations are easily implemented in many quantum systems when compared with multi-qubit operations. We use values of $\tau=1ns$ for the duration of each time interval and a total number of intervals afforded to the optimizing agent of $T=10$. The results are shown in Fig.~\ref{ResultsFigGHZ} where a highest fidelity of $\mathcal{F}(\rho_Q, \sigma) = 0.9746$ is achieved in {$\approx 8$ns}. 
\FloatBarrier
\begin{center}\
	\begin{figure}[t!]\
		\includegraphics[width=0.99\columnwidth]{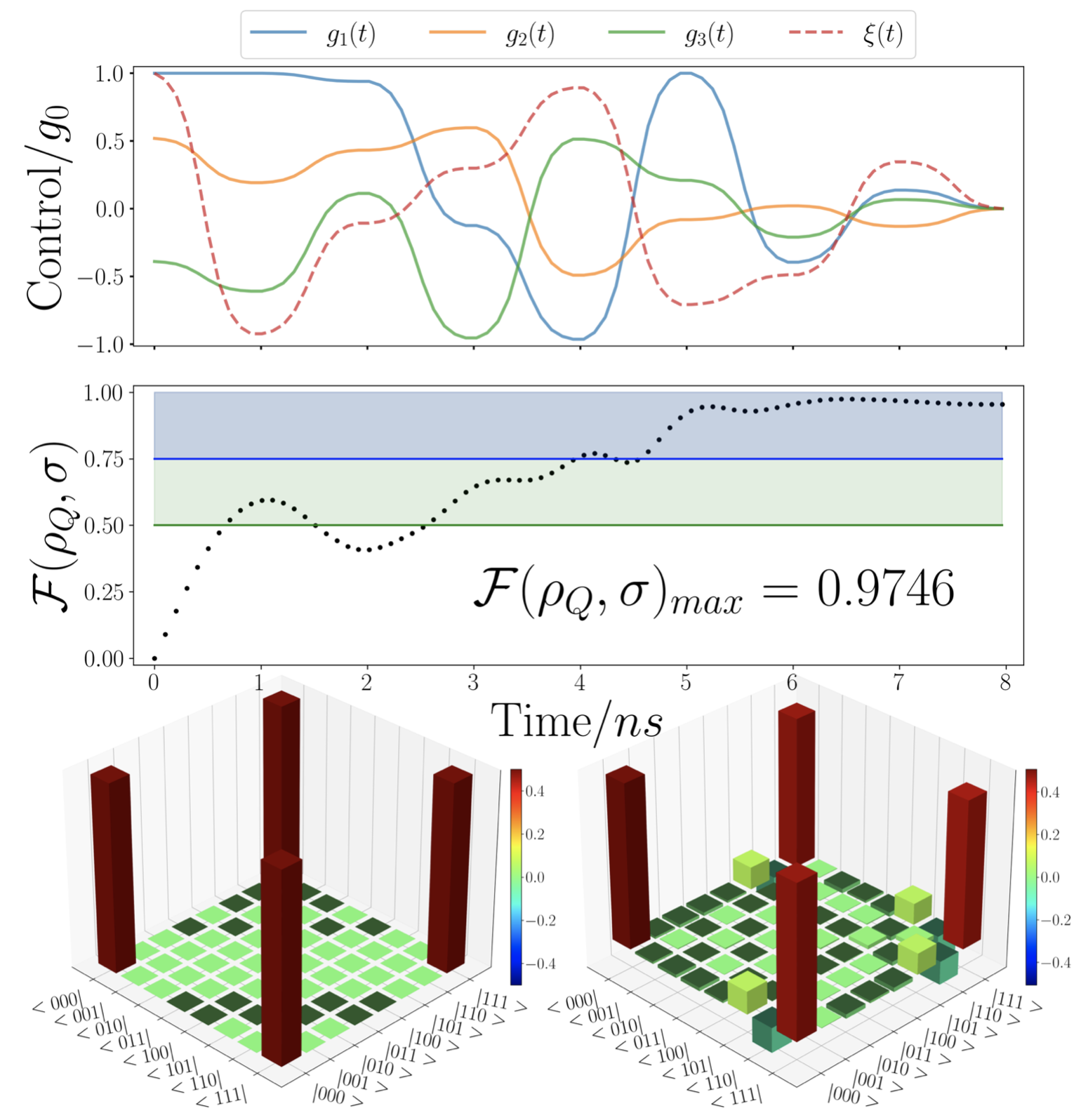}\
		\caption{Results for three-qubit GHZ state. We have the optimal control pulses [Top], the fidelity in the qubit subspace during the dynamics [Middle] and the matrix histogram for the target state (left) and the state with the highest fidelity during the dynamics (right) [Bottom]. Hyper-parameters $\tau=1$ns and $T=10$ are used here, with maximal fidelity achieved, and maintained, within $\approx 8$ns. On the fidelity plot, the horizontal green line is drawn at $c_n = 1/2$ and all fidelities above this exhibit GME (green region). The blue region $c_n = 3/4$ on the other hand shows those fidelities that exhibit both GME and GHZ-class entanglement in particular.} 
			\label{ResultsFigGHZ}
	\end{figure}\
\end{center}\
\FloatBarrier
\paragraph{Three-qubit Dicke state.}
Dicke states embody another class of genuinely tripartite entangled states, inequivalent to GHZ states~\cite{dur2000three}, which have been experimentally realised, projected onto lower dimensional entangled states and employed in open destination teleportation and telecloning~\cite{prevedel2009experimental,chiuri2012experimental}. Specifically we consider the three-qubit two-excitation Dicke state, sometimes referred to as a \textit{flipped W} state, defined as
\begin{equation}
	\ket{{D}^{(2)}_{3}} = \frac{\left(\ket{011} + \ket{101} + \ket{110}\right)}{\sqrt{3}}.
\end{equation}
We assume initial state preparation of $\ket{000}_{Q}\ket{0}_{cav}$, and allow more fine control this time with $\tau=0.5ns$ and $T=20$. 
The maximum fidelity achieved in this instance is $\mathcal{F(\rho_{Q}, \sigma)}_{max} = 0.9896$ within $\approx 7$ns, as shown in figure~\ref{ResultsFigDicke}.
\FloatBarrier
\begin{center}\
	\begin{figure}[t!]\
		\includegraphics[width=0.495\textwidth]{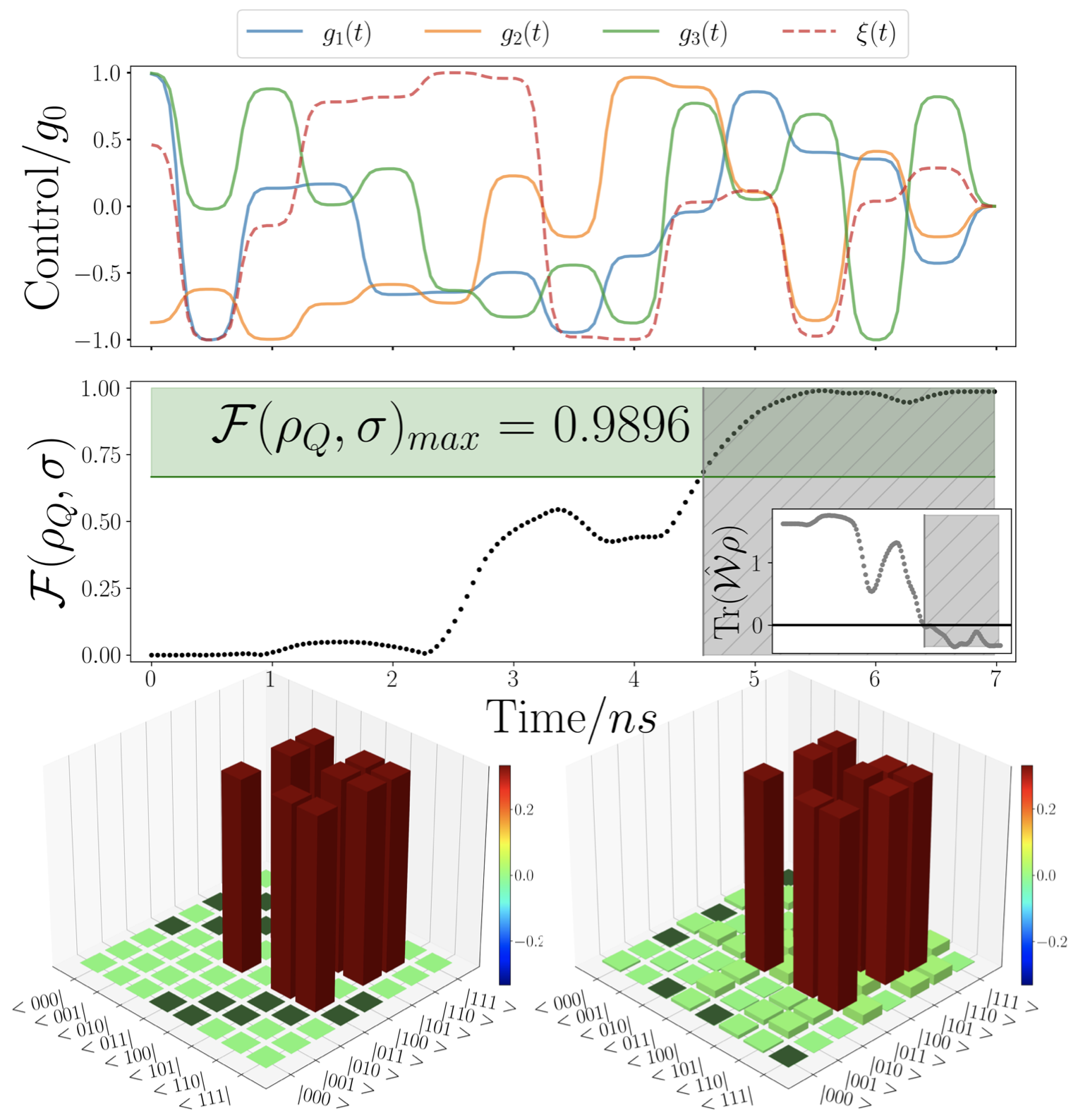}\
		\caption{Results for three-qubit Dicke state. We have the optimal control pulses [Top], the fidelity in the qubit subspace during the dynamics [Middle] and the matrix histogram for the target state (left) and the state with the highest fidelity during the dynamics (right) [Bottom]. The hyper-parameters $\tau=0.5$ns and $T=20$ have been used for these calculations. The corresponding maximal fidelity was achieved within $\approx 7$ns. The green region on the fidelity plot highlights fidelities for which genuine multipartite entanglement (GME) is detected based on the fidelity based witness Eq.~\eqref{fidwit} with $c_n = 2/3$. The inset shows $\text{Tr}(\rho \mathcal{\hat{W}})$ at each instant of time, and the grey region on both the inset plot and the fidelity plot show the temporal region where GME is detected via the collective-spin witness in Eq.~\eqref{spinwit} with $b_s = 3.12$~\cite{toth2007detection}.}\
				\label{ResultsFigDicke}
	\end{figure}\
\end{center}\
\FloatBarrier
\paragraph{Box Cluster State.} Cluster states are a class of graph states useful for measurement based quantum computing~\cite{nielsen2006cluster, bartlett2006simple,walther2005experimental,briegel2009measurement}. Cluster states are represented by graphs of interconnected vertices, where if one starts with each qubit in the ground state, they are defined procedurally by applying a Hadamard gate to each qubit (represented by the vertices of the graph) then applying conditional-phase gates between qubits whose vertices are connected by edges. Here we consider a so-called Box cluster state which is defined by four vertices connected by four edges to form a square. This can be written as,
\begin{equation}
	\ket{\psi_{Box}} = CZ_{41}\left(\Pi^3_{j=1}CZ_{i,i+1}\right)\left(\Pi^4_{j=1}{\cal H}_j\right) \ket{0000}_{1234},
\end{equation}
where $CZ_{i,j}$ is the {controlled-Z} gate between qubits $i$ and $j$ and ${\cal H}_i$ is the Hadamard gate on qubit $i$. Explicitly, the state reads: 
\begin{equation}
	\ket{\psi_{Box}} = \sum_{i,j,k,l=0,1}(-1)^{x_i + x_j + x_k + x_l}\ket{x_i x_j x_k x_l}.
\end{equation}
As before, assuming $\tau = 0.5 ns$ and $T=20$ and starting from the initial vacuum state a maximum fidelity of $\mathcal{F}(\rho_Q, \sigma) = 0.9642$ was achieved (cf. Fig.~\ref{ResultsFigCluster}), this time requiring the full $10ns$ to achieve and maintain maximal fidelity.
\FloatBarrier
\begin{center}\
	\begin{figure}[t!]\
		\includegraphics[width=0.495\textwidth]{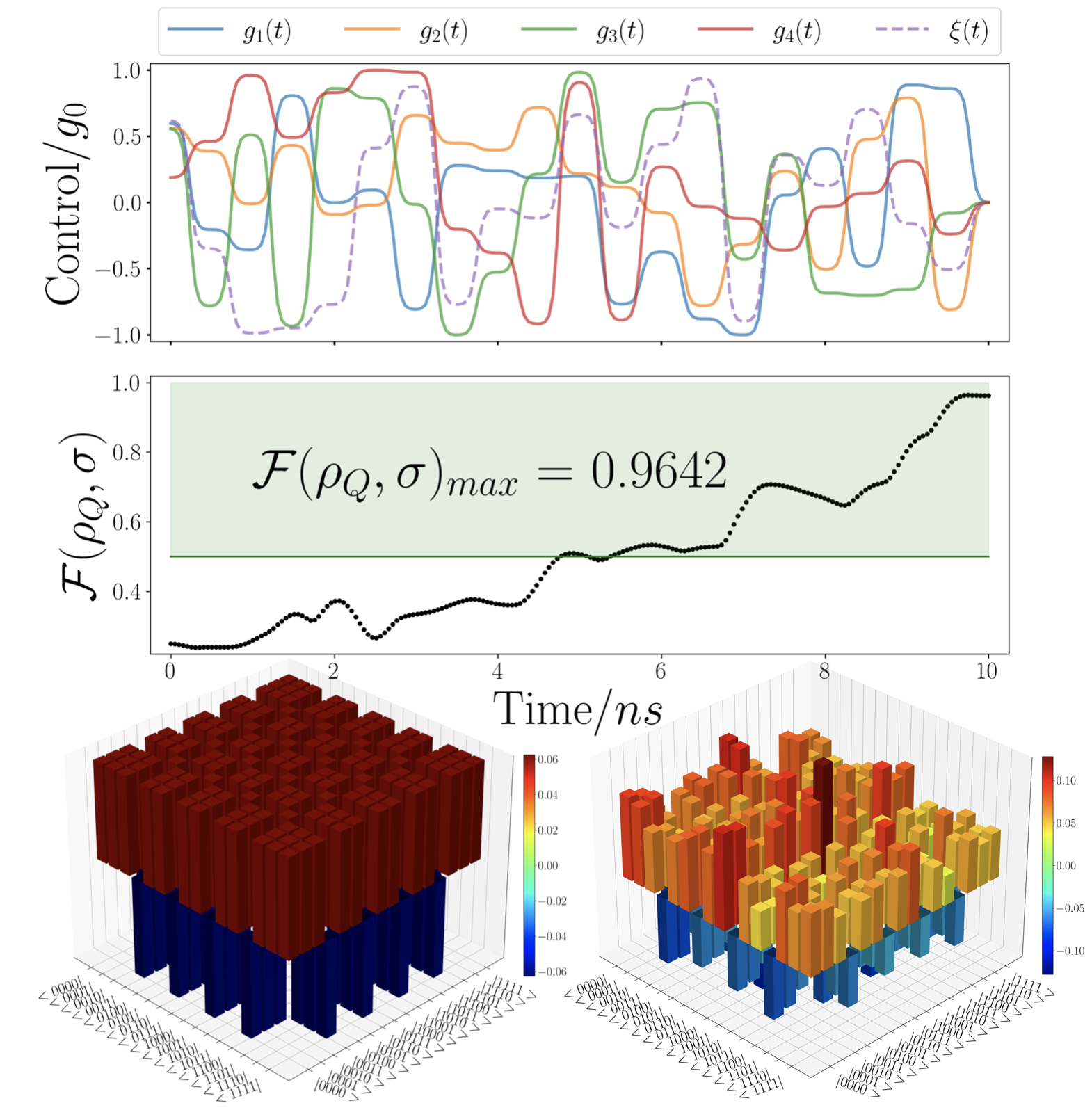}\
		\caption{Results for a four-qubit Box Cluster state. We have plotted the optimal control pulses [Top panel], the fidelity in the qubit subspace during the dynamics [Middle panel] and the matrix histogram for both the target state (left-most figure) and the state with the highest fidelity during the dynamics (right-most one) [Bottom panel]. We have used the hyper-parameters $\tau=0.5$ns and $T=20$ and the maximum fidelity was achieved within $\approx 10$ns. The values of fidelity for which GME is detected via the fidelity based witness Eq.~\eqref{fidwit} with $c_n = 1/2$~\cite{guhne2009entanglement} are shown in the green region.}\
		\label{ResultsFigCluster}
	\end{figure}\
\end{center}\
\FloatBarrier

\subsection{Entanglement Detection}
A central question in any state-engineering scheme is the characterization of the features of the state that has been synthesized. Within the context of our investigation, the core aspect to address is the quantification of multipartite entanglement. 

The task of accurately determining the amount of entanglement in a given quantum state, which is challenging in general, 
is made even more difficult in multipartite settings 
due to the hierarchical structure of entanglement in many-body systems~\cite{horodecki2009quantum, plenio2014introduction, christandl2004squashed, guo2020multipartite, szalay2015multipartite} 
and the need to construct  convex-roof extensions of any pure-state quantifier, when dealing with mixed states~\cite{gao2008entanglement, cao2010efficient, rothlisberger2009numerical}. 

A significant tool in these endeavours is embodied by  entanglement witnesses, which often offer experimentally viable ways of detecting (or even quantifying~\cite{brandaoquantification}) entanglement~\cite{chruscinski2014entanglement, guhne2009entanglement} that are suitable for mixed states and have been used to detect genuine multipartite entanglement (GME) already for GHZ class, W-class and graph states~\cite{toth2007detection, campbell2009characterizing,jungnitsch2011entanglement}. Whilst high fidelity with a maximally entangled state is a good indicator that we have generated entanglement close to the right {\it structure}, it is useful to quantitatively check for GME in the system. A natural approach is to use so-called ``fidelity based" entanglement witnesses. These are of the general form 
\begin{equation}
\label{fidwit}
    \mathcal{\hat{W}}_{F} = c_n \mathds{1} - \op{\psi}{\psi},
\end{equation}
where $\ket{\psi}$ is the state of interest and $c_n$ is the maximal overlap between $\ket{\psi}$ and all bi-separable states. Thus, any state for which Tr$(\rho \hat{\mathcal{W}}_{F}) \geq 0$ is bi-separable and consequently Tr$(\rho \hat{\mathcal{W}}_{F}) < 0$ indicates genuine multipartite entanglement. The overlap values $c_n$ have already been calculated for 3-qubit GHZ and W states, and 4-qubit linear cluster states~\cite{toth2007detection,guhne2009entanglement}, which up to local unitaries and swaps coincide exactly with the three cases considered above. In light of the analysis reported in Figs.~\ref{ResultsFigGHZ},~\ref{ResultsFigDicke}, and~\ref{ResultsFigCluster} these witnesses are readily implemented. Considering $\text{Tr}(\rho \mathcal{\hat{W}}_{F}) = c_n\text{Tr}(\rho) - \text{Tr}(\rho \op{\psi}{\psi}) < 0$, where of course $\text{Tr}(\rho) = 1$ and $\text{Tr}(\rho \op{\psi}{\psi}) = \expval{\psi}{\rho}$ is the fidelity between $\rho$ and $\ket{\psi}$. This is clearly fulfilled when $\mathcal{F}(\rho, \ket{\psi}) > c_n$ so we can simply highlight the threshold value ($c_n$) of fidelity above which GME can be detected. 
The region where this happens is highlighted in green in each of the figures.

One may also be interested in not only ensuring that the state has GME but also, in the GHZ case, if we have generated GHZ-class entanglement~\cite{acin2001classification}. In this case, $c_n$ will be the maximal overlap between $\ket{GHZ}$ and all W-class entangled states. This region is highlighted in blue in Fig.~\ref{ResultsFigGHZ}.

Finally, other constructions of witnesses are useful for specific states; for example, witnesses based on collective spin operators have been used to more efficiently detect GME in symmetric Dicke states as in~\cite{toth2007detection}. Here the witness takes the form
\begin{equation}
    \label{spinwit}
    \hat{\mathcal{W}} = b_s \mathds{1} - (\hat{J}_x^{2} + \hat{J}_{y}^{2}),
\end{equation}
where $\hat{J}_{k}$ is the collective $k=x,y$ spin operator~\cite{campbell2009characterizing}. It is not so straightforward here to place a delineation at a given fidelity as with the fidelity based witnesses, so we first plot $\text{Tr}(\rho \mathcal{\hat{W}})$ as an inset within figure~\ref{ResultsFigDicke} and highlight the region for which $\text{Tr}(\rho \mathcal{\hat{W}}) < 0$. This region is then shown as the grey hatched area on the larger fidelity plot. It can be seen that both the fidelity based witness and the collective spin based witness detect GME at strikingly similar times.

\subsection{Decoherence}
\label{decoherence}
\begin{figure}
	\includegraphics[width=1.1\linewidth]{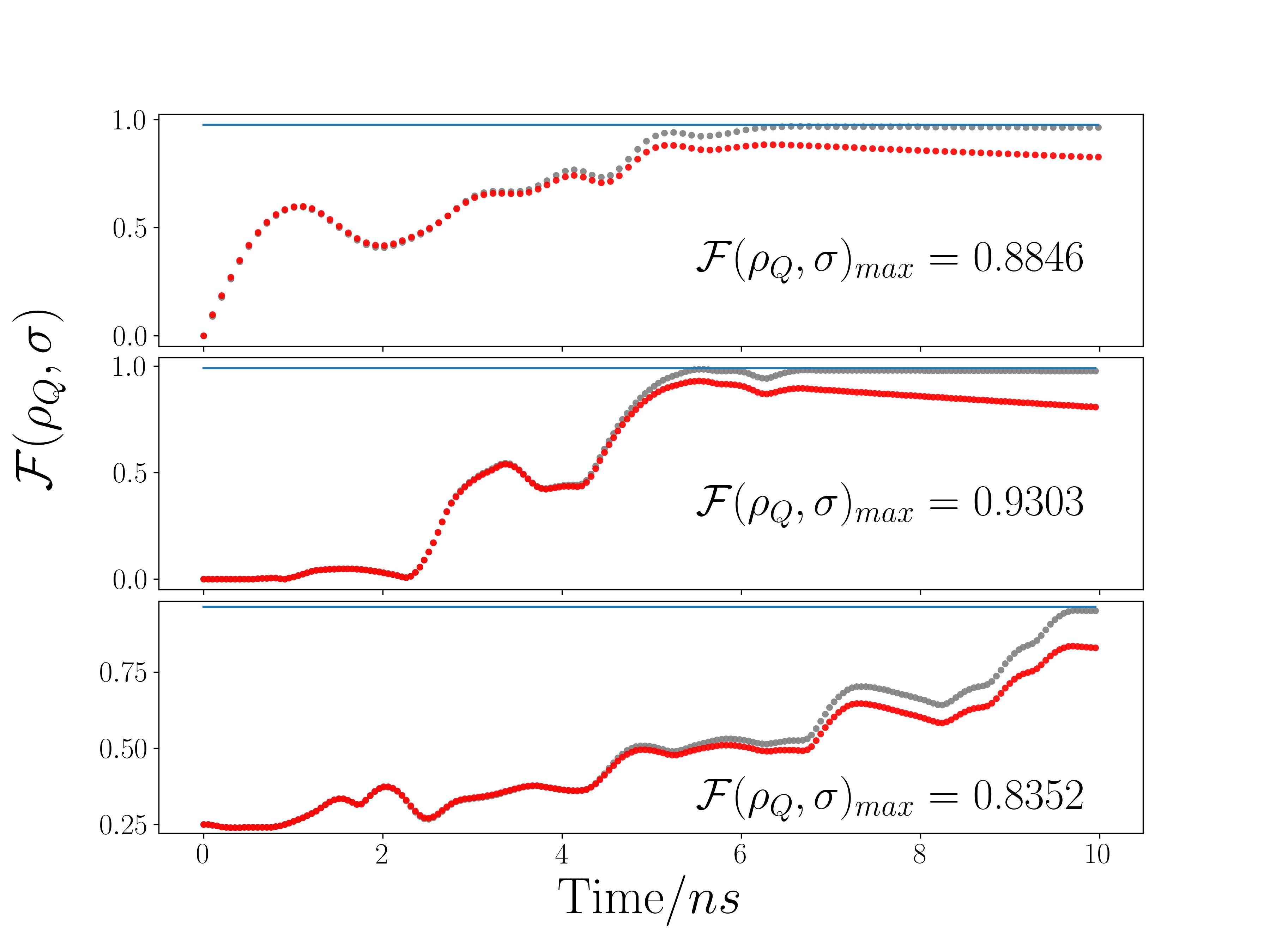}
	\caption{A comparison between the fidelity achieved for the (Top) GHZ state preparation, (Middle) Dicke state preparation and (Bottom) Box Cluster state preparation schemes in the presence of cavity decay and qubit dephasing only (Grey) and qubit decay only (Red). Each plot shows the maximum achieved fidelity in the latter case. The horizontal blue line shows the maximum fidelity achieved in the ideal, noiseless case.}
		\label{DecoherenceFid}
\end{figure}

So far, we have exclusively considered closed system dynamics and as such the optimality of the control schemes presented is limited to the noiseless case. It is of interest then to assess how these control schemes perform in the presence of decoherence. Specifically we can write the following Lindblad master equation to model the effect of decay and dephasing acting on each of the constituent subsystems~\cite{breuer2002theory, manzano2020a}

\begin{equation}
	\dot{\rho} = -i\comm{\tilde{H}}{\rho} + \kappa \mathcal{D}\left[a\right]\rho + \sum_{j}^{N} \left( \gamma_j \mathcal{D} \left[ \sigma^{-}_{j} \right] \rho +2\gamma_{\phi, j} \mathcal{D} \left[ \sigma^{+}_{j} \sigma^{-}_{j} \right] \rho \right),
\end{equation}
where $\kappa$ is the cavity decay rate, $\gamma_{\phi,j}$ and $\gamma_j$ are the dephasing and decay rate for the $j^{th}$ qubit, respectively, and we have introduced the superoperators 
\begin{equation}
    {\cal D}[Q]\rho= Q\rho Q^{\dagger} - \frac{1}{2}\left\{Q^{\dagger} Q, \rho\right\}
\end{equation}
for an arbitrary operator $Q$. If we adopt physically reasonable values for these rates we can obtain an estimate of the performance of a physical system. For example, considering superconducting systems we set $\kappa = 2\pi \times 5 \text{ kHz}$ for the cavity damping and the typical values of $2 \pi \times 300 \text{ KHz}$ and $2\pi \times 5 \text{ MHz}$ for the dephasing and decay rate for each of the qubits~\cite{majer2007,blais2020circuit}. In these systems, the use of high-Q cavities and low temperatures leads to a reduced cavity decay and dephasing rate. The qubit decay is thus the main source of decoherence. In these conditions, the effects of noise is reported in Fig.~\ref{DecoherenceFid}. We can clearly see that the control protocols are almost completely insensitive to cavity decay and qubit dephasing, whilst still reasonably robust against qubit decay.

\section{Conclusions}
\label{conc} 

We have investigated the use of evolutionary algorithms, which are well-established strategies for both classical and quantum control, for direct quantum state engineering and multipartite entanglement generation. Specifically, we considered the effective Hamiltonian of a network of non-interacting qubits jointly addressed by a common driven bus and applied a genetic algorithm to identify the set of optimal pulses to drive the evolution of the qubit register. This has allowed us to successfully put forward robust protocols for the engineering of three- and four-qubit states that play a crucial role in quantum metrology and computing, including Dicke and cluster states. The protocols, which offer significant robustness to the most common and crucial sources of imperfection, provide further evidence of the benefit of a hybrid approach to quantum control that puts together the insight provided by machine learning strategies to well-established schemes for optimal control. The extension of these approaches to larger registers and non-unitary dynamics will pave the way to quantum process engineering enhanced by machine learning and optimised by quantum control methods.  
\FloatBarrier
\acknowledgements

We acknowledge support from the European Union's Horizon 2020 FET-Open project TEQ (766900), the Horizon Europe EIC Pathfinder project QuCoM (Grant Agreement No. 101046973), the Leverhulme Trust Research Project Grant UltraQuTe (grant RGP-2018-266), the Royal Society Wolfson Fellowship (RSWF/R3/183013), the UK EPSRC (grant EP/T028424/1) and the Department for the Economy Northern Ireland under the US-Ireland R\&D Partnership Programme.

\FloatBarrier

\appendix
\section{Functional Form for controls}
\label{app1}
\begin{figure}
    \label{tanhDetails}
	\begin{minipage}{0.8\linewidth}
		\includegraphics[width=\linewidth]{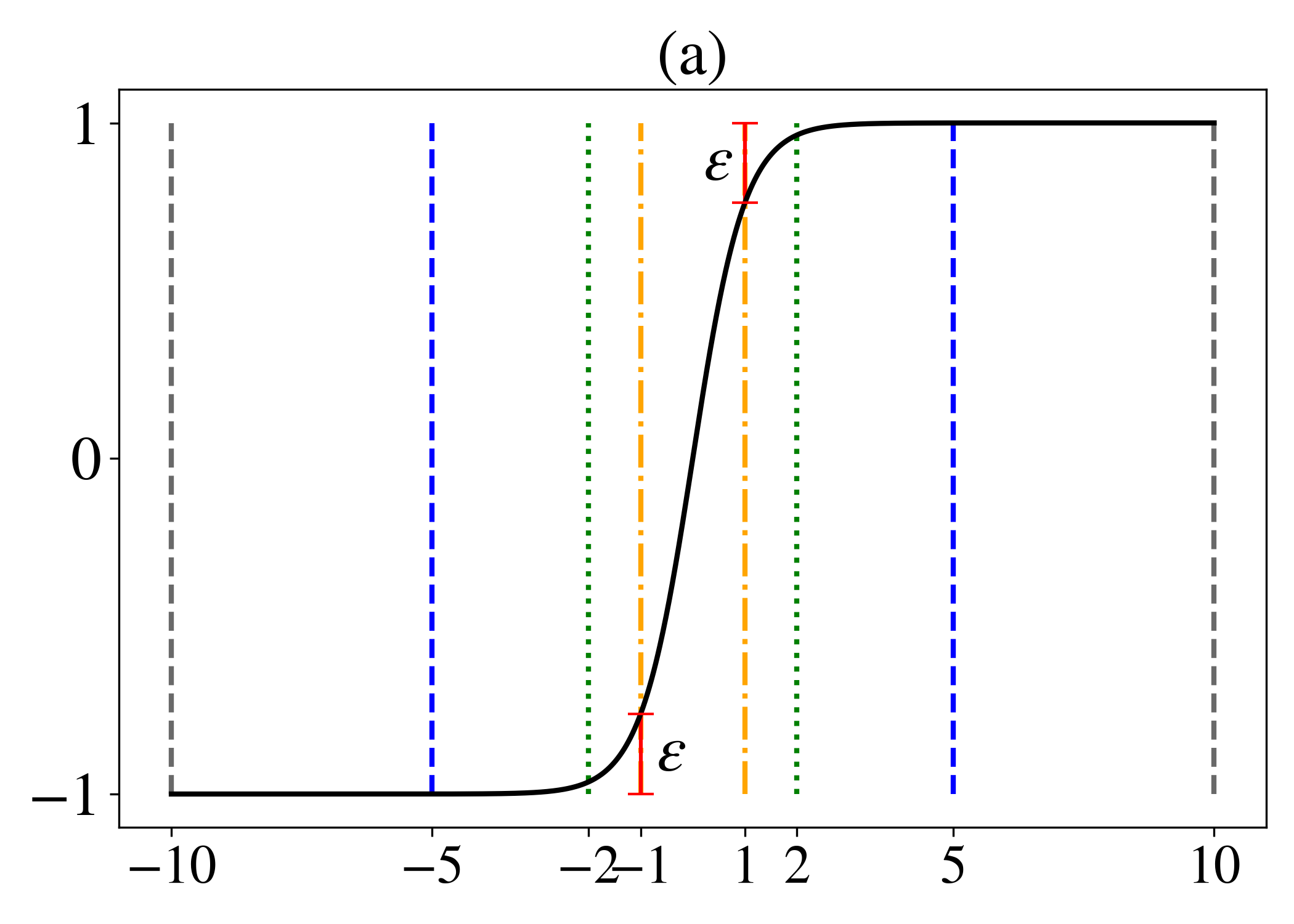}
	\end{minipage}
	\begin{minipage}{0.8\linewidth}
		\includegraphics[width=\linewidth]{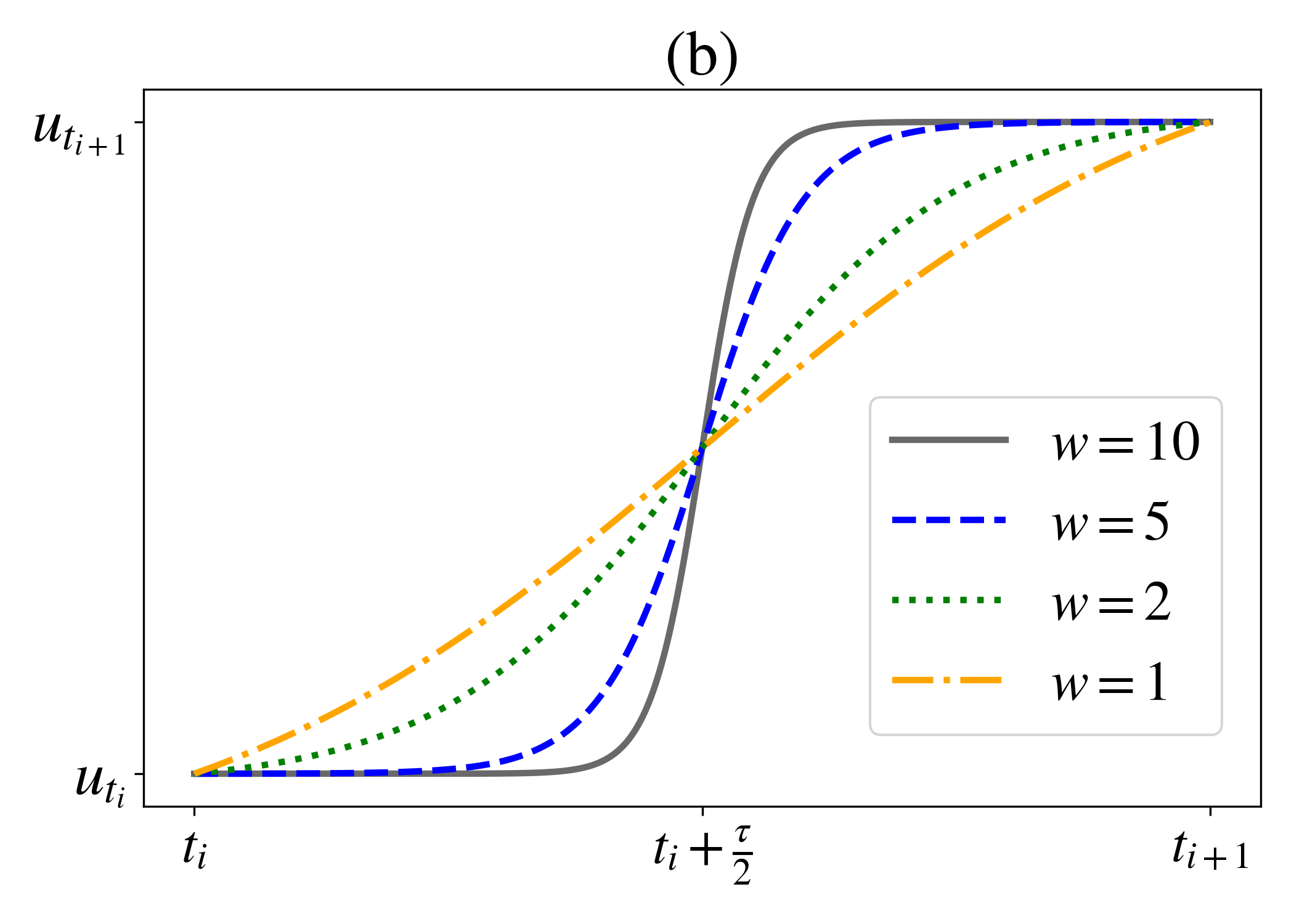}
	\end{minipage}
	\caption{Comparison of how the \textit{clipping window} affects the shape of the interconnecting tanh functions used to construct the control functions. In (a) the tanh function is shown with over-laid dashed lines representing different widths of this clipping window. From (a) we can see how the clipping window width determines the severity of the time dependence. Namely, taking the largest width (grey) leads to a more ``step-like" dependence, as evidenced in (b) where each of the clipped functions are scaled into a time interval of equal duration $\tau$ and re-scaled to account for the error $\varepsilon$. (a) shows how the error $\varepsilon$ increases as the clipping window narrows, i.e the narrowest (yellow) window has the largest error thus requiring the most re-scaling.}
\end{figure}

A common approach to optimal control problems is to break the dynamics up into $T$ time intervals of equal duration $\tau$, then at each interval assign a constant value to each control parameter. This results in piece-wise constant (PWC) control ``pulses", $u(t) \to \{u_i\}_{i=0,...T-1}$, where $u$ is routinely used to denote a generic control function. Such discretization is useful for application of Reinforcement Learning techniques as in~\cite{porotti2019coherent, paparelle2020digitally, sgroi2020reinforcement, brown2021reinforcement, giannelli2022tutorial}. Whilst useful, this type of functional form includes discontinuities, often in the form of instantaneous jumps in the control, which is experimentally unfeasible and subsequently requires some method of smoothing post-optimization often to the detriment of performance. Here we use a generalised form of this discretisation, where instead of allowing the optimization to chose the $T$ constant values corresponding to the control value during each time interval, we allow it to chose the $T+1$ values corresponding to the start(/end) of each interval, then connect each value with a smooth, time dependent function during the interval.  Namely, the functional time dependence of each control pulse is made to be a clipped $\tanh$ function centred in the middle of the time interval. For example consider a simple $\tanh$ function, shown in figure A. If we clip the $\tanh$ function within windows of different widths centred around zero we can make a ``step-like" like function with varying severity. This is tantamount to scaling the $\tanh$ function along the y-axis and clipping it at $\pm 1$.  So the time dependence within time interval $i$, $t_i \leq t \leq t_{i+1}$ can be written as,

\begin{equation}
	f_i(t) = \begin{cases} 0 &\mbox{if } t < t_i \\
		(u_{i+1} - u_{i})  \left(\frac{S \tanh(Wt - (t_i + \frac{\tau}{2})) + 1}{2}\right) + u_{i} &\mbox{if } t_i \leq t < t_{i+1} \\
		0 &\mbox{if } t \geq t_{i+1}
	\end{cases}
\end{equation}
where $u_i$, $u_{i+1}$ is the value of the control at times $t_i$, $t_{i+1}$ respectively, $W$ determines the severity of the step, and $S$ is a scaling factor introduced to deal with the error $\varepsilon$, as in fig~\ref{tanhDetails}.
Therefore the complete functional for a general control, under this scheme, is given by
\begin{equation}
    f(t) = \sum_{i = 0}^{T} f_i(t),
\end{equation}
where, again, $T$ is the number of time intervals. Full functional forms can be clearly seen in figures~\ref{ResultsFigGHZ},~\ref{ResultsFigDicke}, and \ref{ResultsFigCluster}.
\section{Algorithm Implementation Details.}
\label{algo_details}
 In each case the population size for the algorithm was $\mathcal{N}_{pop} = 48$ and was determined by the number of CPUs available, since each chromosome in the population was evaluated in parallel to significantly speed up computation time. The number of survivors was fixed at $\mathcal{N}_{pop}/2 =24$, from which $\mathcal{N}_{pop}/4 = 12$ pairs of parents were selected according to a probability distribution determined by their relative fitness. Each pair of parent chromosomes produced 2 offspring chromosomes to re-populate. The mating procedure involved 2 steps. After making a one copy of each parent chromosomes: 1) Each separate section of $T+1$ elements - corresponding to the different control pulses - were completely swapped between the two copied chromosomes with $\approx 50\%$ probability, otherwise they were left unchanged. 2) random indices were then selected and the combination given by equation~\ref{combination} was applied, where $\beta$ was randomly sampled from [0,1] for each separate index, in an inverse manner to the two copies. This results in 2 new offspring chromosomes that are completely complementary to one and other. Mutation was then applied by selecting chromosomes (apart from the fittest) at random and random indices within these chromosomes to replace with completely random values. The rate $\alpha$ determined the total number of parameter values within the entire generation that were flipped and generally assumed a value of $\alpha\approx0.2$.

\section{Simulation Details}
\label{simulation_details}
Simulations were carried out using the Numerical Schrödinger Equation solver within the QuTiP package in python~\cite{johansson2012qutip}. Thus, for the simulations we had to approximate the harmonic mode by a $d$-dimensional harmonic system, however care must be taken. Since the cavity is assumed to be resonantly driven, if we use a low value of $d$ and/or too strong a drive, then the $d$-dimensional harmonic system will quickly become saturated, and in fact early optimizations used this fact to produce extremely good results with low dimensional cavities. Of course when one then simulated these controls with higher levels in the cavity the performance was destroyed and so the results were neither realistic nor practically useful. One can avoid this in two ways: 1) By encoding enough redundancy in the cavity by using very high values of $d$, which increases computational cost; or 2) Including a term in the numerical fitness function that punishes population of the higher energy states of the cavity approximation. Here we apply both by choosing $d \in [5,6]$ as well as including the term 
\begin{equation}
 \phi_1=   - \frac{\nu}{\tau T} \int_{0}^{\tau T} \expval{n}{\rho_{cav}(t)} dt,
\end{equation}
where $\ket{n}$ is the highest excited state within the $d$-dimensional approximation and $\rho_{cav}$ is the reduced state of the cavity. $\nu$ determines the strength of punishment and was set at $\nu=0.1$. (Practically, since the dynamics is solved numerically the integral was actually a summation).

Another issue one encounters with this type of simulation is that, if we assess fitness based on the outright maximum value of the fidelity during the induced dynamics, then it is possible to observe successful control schemes that induce sharp spikes in the fidelity landscape. If the control scheme induces such spikes on a time scales shorter than that required to completely ``switch off" all of the controls then it is impossible to extract the state of maximum fidelity and the control sequences again cease to be practically useful. We can combat this by including an additional term in the fitness function that rewards control schemes that briefly maintain near maximal fidelity for a short time, allowing us to selectively uncouple the cavity whilst maintaining the state of maximal fidelity in the qubit subspace. The term used was
\begin{equation}
  \phi_2=  + \frac{\mu}{m\tau} \int_{t_{max}}^{t_{max} + m\tau} \mathcal{F}(\rho_{Q}(t), \sigma) dt,
\end{equation}
where $m$ is the number of time intervals of length $\tau$ to include in the numerical bonus beyond which we no longer care if the fidelity deteriorates. $\mu$ is again a variable that determines the relative importance of maintaining fidelity after maximum and was set to $\mu=0.5$. Thus the actual reward function employed was
\begin{align}
    \textit{Chromosome fitness} =& \left[ \mathcal{F}(\rho_{Q}(t_{max}), \sigma)\right]+\phi_1+\phi_2 
\end{align}
where $t_{max}$ is the time at which maximum fidelity is achieved during the induced dynamics. This leads to control schemes that maximise and briefly maintain fidelity allowing us to selective switch of the couplings, whilst also only exclusively utilising lower lying levels of the harmonic mode. Thus in principle the resulting controls could be practically implemented and yield identical performance.
Clearly, these considerations are necessitated by the use of simulation and the case is much simpler if one wishes to use the algorithm on a physical system, however in this case the ability to parallelise the computational steps, one major advantage of the Genetic Algorithm, is suppressed.

\FloatBarrier

\end{document}